\documentclass[epj]{svjour}
\usepackage{graphics}
\newcommand{\beq}{\begin{equation}}
\newcommand{\eeq}{\end{equation}}
\def\eq#1{{(\ref{#1})}}

\newcommand{\as}{\alpha_s}

\begin{document}
\title{Parton energy loss at strong coupling and the universal bound}
%\subtitle{The universal bound}
\author{Dmitri E. Kharzeev\inst{1} 
}                     % Do not remove
\institute{Nuclear Theory Group, Physics Department, 
Brookhaven National Laboratory, Upton, NY 11973, USA}
\date{Received: date / Revised version: date}
\abstract{
The apparent universality of jet quenching observed in heavy ion collisions at RHIC for light and heavy quarks, as well as for quarks and gluons, is very puzzling and calls for a theoretical explanation.   
Recently it has been proposed that the synchrotron--like radiation at strong coupling gives rise 
to a universal bound on the energy of a parton escaping from the medium. Since this bound appears quite low, almost all of the observed particles at high transverse momentum have to originate from the surface of the hot fireball. Here I make a first attempt of checking this scenario against the RHIC data and formulate a "Universal Bound Model" of jet quenching that can be further tested at RHIC and LHC.
 % end of PACS codes
} %end of abstract
\maketitle
\section{Introduction}
\label{intro}
RHIC experiments have observed a striking deficit of the
high transverse momentum particles \cite{Arsene:2004fa,Adcox:2004mh,Back:2004je,Adams:2005dq}: 
their yield is far smaller than one would expect from an incoherent superposition of nucleon--nucleon collisions. Such a suppression resulting from the parton energy loss has been predicted  \cite{Bjorken:1982tu,Gyulassy:1993hr,Baier:1996kr} as a signature of the formation of dense quark-gluon matter. Since we are dealing with hard processes, it seems natural to start the analysis of this phenomenon basing on QCD perturbation theory. In this approach, the leading mechanism of energy loss for an ultra-relativistic parton is the induced radiation of gluons. Indeed, 
the radiative energy loss \cite{Gyulassy:1993hr,Baier:1996kr} has been found to describe the magnitude of the observed suppression, see for example reviews  \cite{Baier:2000mf,Gyulassy:2003mc}. 

Radiative energy loss mechanism can be definitively tested by using heavy quarks \cite{Dokshitzer:2001zm} -- since heavy quarks move with velocity $v < c$, the induced radiation must be depleted due to the "dead cone" effect -- the vanishing of radiation intensity in the forward direction. This reduces the amount of energy loss, and results in weaker suppression for charm and bottom quarks. 
The heavy-to-light ratios at high transverse momentum have thus been predicted to exceed unity \cite{Dokshitzer:2001zm}; this conclusion survives after the differences in the production mechanisms and fragmentation functions for heavy and light quarks are taken into account \cite{Armesto:2005iq,Djordjevic:2004nq}. 

It thus came as a surprise when RHIC experiments \cite{Abelev:2006db,Adare:2006nq} observed a strong suppression of the high transverse momentum electrons originating from the decays of charmed and beauty hadrons. The magnitude of the observed suppression indicates that heavy and light quarks are attenuated very similarly in hot QCD matter, in sharp disagreement with the theoretical expectations. This remains true even in the range of transverse momenta that may be dominated by the decays of $b$ quarks \cite{Abelev:2006db,Adare:2006nq}. Moreover, the momenta of the heavy quarks also seem to be strongly deflected by the medium, as is evidenced by the observed elliptical flow (azimuthal anisotropy of the produced heavy quarks with respect to the reaction plane) \cite{Adare:2006nq,:2008hj,Awes:2008qi}.
While it is not yet entirely clear that a perturbative approach cannot be reconciled with the data, the heavy quark energy loss puzzle as well as some other observations in high transverse momenta phenomena \cite{Arsene:2004fa,Adcox:2004mh,Back:2004je,Adams:2005dq} provide ample  motivation to think about alternatives. 

One such alternative is to consider parton propagation in coherent external fields, rather than multiple scattering on static sources. Another 
possibility is to look at non--perturbative effects. Both of these alternatives are related at strong coupling: indeed, in this case the well-defined quasi-particles which would play the role of static scattering centers do not exist, and the non-perturbative effects dominate.

Unfortunately, the set of tools that can be used to address the real--time dynamics of QCD in the strong coupling domain is limited. 
Therefore any information about the behavior of gauge theories at strong coupling is very valuable and may provide a hint on how to deal with non-perturbative effects in QCD matter.

\medskip

There has been a major breakthrough in the understanding of strong coupling dynamics of ${\cal N}=4$ SUSY Yang-Mills theory made possible by the AdS/CFT correspondence \cite{Maldacena:1997re,Gubser:1998bc,Witten:1998qj,Aharony:1999ti}. 
In particular, the strong coupling dynamics on the gauge theory side appears dual to the classical supergravity in $AdS_5 \times S_5$ space. 
Of course, QCD and ${\cal N}=4$ SUSY Yang-Mills are very different theories, and conformal invariance of the latter is a crucial 
property which determines the metric of the $AdS_5$ space. Conformal invariance results in the absence of confinement and asymptotic freedom in  ${\cal N}=4$ SUSY Yang-Mills theory. 
Thermodynamical and transport properties of ${\cal N}=4$ SUSY Yang-Mills are also different; in particular the bulk viscosity (related to the scale anomaly of QCD \cite{Kharzeev:2007wb,Karsch:2007jc,Meyer:2007dy}) vanishes unless the conformal symmetry is broken  \cite{Benincasa:2005iv,Buchel:2007mf,Gubser:2008yx}. Nevertheless one hopes that at least some aspects of the plasma behavior at strong coupling may be universal for ${\cal N}=4$ SUSY Yang-Mills and QCD at temperatures higher  than the deconfinement and chiral symmetry restoration temperature (but not much higher so that the coupling is still strong). 
Various properties of the plasma have been considered through the AdS/CFT correspondence, including the computation of  the shear viscosity-to-entropy ratio in the strong coupling limit \cite{Policastro:2001yc,Kovtun:2004de}. The Langevin dynamics of drag force acting on a massive quark traversing the ${\cal N}=4$ SUSY Yang-Mills plasma has been determined by solving the dual problem of string trailing in the ${\rm AdS_5}$ Schwarzschild background \cite{Herzog:2006gh,Gubser:2006bz,CasalderreySolana:2006rq}; a comparison of AdS/CFT drag and pQCD predictions for observables in heavy ion collisions has been recently performed in \cite{Horowitz:2008ig}. A different approach proposed in \cite{Liu:2006ug} aims at matching ${\cal N}=4$ SUSY Yang-Mills calculations to the perturbative QCD. The related problem has been discussed in a number of papers, including \cite{Chernicoff:2006hi,Argyres:2006yz,Dominguez:2008vd,Hatta:2008tx,Mueller:2008zt}.

\medskip

In this talk I will first describe the result of Ref. \cite{Kharzeev:2008qr}, then formulate a simple model 
and test it against some of the available RHIC data. Readers familiar with \cite{Kharzeev:2008qr} can proceed directly to Sec.  \ref{ubm}.

\section{Parton energy loss at strong coupling}

The problem of energy loss by an accelerating external source in ${\cal N}=4$ SUSY Yang-Mills theory in the large $N$ limit  has been addressed by Mikhailov \cite{Mikhailov:2003er}. He found that the energy loss is given by the following formula:
\beq\label{lien}
E_{rad}(T) = \frac{\sqrt{\lambda}}{2 \pi} \int_{- \infty}^T \ d t \ \frac{\vec a^2 - (\vec v \times \vec a)^2}{(1 - \vec v^2)^3},
\eeq
where $\lambda=g^2 N$ is 't Hooft coupling, $\vec v$ is 
velocity and $\vec a = \dot{\vec v}$ is acceleration of the charge; we put $c = 1$. Remarkably, with the substitution 
\beq \label{subst}
\frac{\sqrt{\lambda}}{2 \pi} \leftrightarrow \frac{2 e^2}{3}
\eeq
this is exactly the Li\'enard formula \cite{lienard} dating back to 1898 which describes the radiative energy loss in external electromagnetic fields in classical electrodynamics \cite{Landau,Jackson}. In the case of electrodynamics, the linear dependence of the radiated energy on the path length is a consequence of the linearity of Maxwell equations: the emission of electromagnetic radiation at any time is not affected by the previously emitted radiation.  

Yang-Mills equations however are non-linear, and one expects, and indeed finds at weak coupling \cite{Baier:1996kr}, a non-linear dependence of the radiated energy on the path length.  This non-linear dependence is a purely non-Abelian effect: the amplitude of radiation at any given time  is affected by the previously emitted radiation.  The linear local dependence of the energy lost by the quark in the strong coupling regime as given by \eq{lien} is thus a highly non-trivial result. 

The method of \cite{Mikhailov:2003er} is based on considering a Wilson loop with boundaries given by the external quark and anti-quark sources; in $AdS_5\times S_5$ space, 
the quark and anti-quark are connected by a classical string with two boundaries. The worldsheet of this string is an extremal surface in $AdS_2 \subset AdS_5$, and the energy of the accelerating quark is lost to the excitations on this surface -- non-linear waves. The propagation  of the nonlinear 
wave on the string worldsheet in the large $N$ limit is described by  
the classical sigma model \cite{Mikhailov:2003er}, and the linear dependence of the final result \eq{lien} on the path stems from the integrability of this model \cite{Mandal:2002fs,Bena:2003wd}\footnote{If it appears that this linear Abelian-like formula holds in QCD at strong coupling, the reason may be the conjectured quasi-Abelian dominance proposed by 't Hooft \cite{'t Hooft:1981ht}; this would imply a dynamical role for magnetic monopoles, also in the plasma \cite{Korthals Altes:2006gx,Shuryak:2007qs}.}.  In the large $N$ limit the interactions with closed strings are suppressed, and so the entire lost energy can be attributed to the non--linear wave on the extremal surface stretched between the quark and anti-quark. It is yet unclear (at least to the present author) what excitation corresponds to this wave in the dual Minkowski space gauge theory language; nevertheless below we will attempt to give a qualitative picture based on the analogy with electrodynamics of strong fields.   

\medskip

It should be noted that once the linear, local dependence on the path is established, Lorentz  invariance and dimensional counting completely determine the structure of the formula for the lost energy \cite{Landau,Jackson} -- it has to be proportional to the Li\'enard formula in classical electrodynamics.
Indeed, the relativistic expression for the energy-momentum vector $P^{\mu}$ of the emitted radiation  reads  \cite{Landau,Jackson}, with the substitution \eq{subst}
\beq\label{relat}
P^{\mu} = \frac{\sqrt{\lambda}}{2 \pi} \ \int \frac{d^2 x_\rho}{d \tau^2}\  \frac{d^2 x^\rho}{d \tau^2} \ d x^{\mu},
\eeq
where $\tau$ is the proper time; this is exactly Mikhailov's result \cite{Mikhailov:2003er}. 
 This means that the coincidence of the result \eq{lien} with the Li\'enard formula cannot be considered as an evidence that the mechanism of energy loss at strong coupling is classical radiation; having this in mind, we will nevertheless for simplicity use the familiar language of electrodynamics and refer to the flow of the emitted energy and momentum described by \eq{lien} and \eq{relat} as "radiation". 

The result of Mikhailov \cite{Mikhailov:2003er} has been verified and extended by 
Sin and Zahed \cite{Sin:2004yx} and by Chernicoff and Guijosa \cite{Chernicoff:2008sa}. Sin and Zahed in particular have argued that high momentum partons in a strongly coupled plasma would not be able to penetrate beyond the distance of $1/\pi T$ \cite{Sin:2004yx}. Chernicoff and Guijosa have derived the expression for the dispersion relation of moving quark, and have considered also the case of finite temperature  \cite{Chernicoff:2008sa}.
 
\medskip  
 
We will see that under quite natural assumptions about the evolution of gauge fields in heavy ion collisions, the result \eq{lien} implies the existence of a universal (i.e. independent of the initial energy, but dependent on the mass of the parton and on centrality of the collision) upper bound on the final energy of the parton escaping from the strongly coupled matter \cite{Kharzeev:2008qr}. 
 
 Let us begin by considering the special case of acceleration parallel to the velocity, $\vec a \parallel \vec v$. Introducing 
 $\gamma = 1/\sqrt{1-v^2}$ and the momentum $\vec p = \gamma m \vec v$,  we get from \eq{lien} or \eq{relat} 
the expression for the power of radiation (radiated energy per unit time):
 \beq
\frac{d E_{rad}}{d t} = \frac{\sqrt{\lambda}}{2 \pi}  \ \frac{1}{\gamma^2 m^2}\ \left(\frac{d \vec p}{d \tau}\right)^2,
\eeq
Since $d \tau = d t / \gamma$, and $d \vec p / d \tau = \gamma\ d \vec p / d t = \gamma \vec F$, where $\vec F$ is the force acting on the charge we get 
 \beq\label{radpar}
\frac{d E_{rad}}{d t} = \frac{\sqrt{\lambda}}{2 \pi} \frac{1}{m^2}\ \ \vec F^2.
\eeq 
The radiation power $d E_{rad}/d t$ in the case of $\vec a \parallel \vec v$ is thus independent of the energy of the charge and is determined only by the magnitude of the external force, as is well known from
 classical electrodynamics \cite{Landau,Jackson}. The situation when $\vec a \parallel \vec v$ is encountered for example when a quark jet propagates in the vacuum and is slowed down by the force of the string $F = dp/dt = dE/dx = \sigma$, where $\sigma$ is the string tension. In this case the rate of quark energy loss as given by \eq{radpar} does not depend on energy.
 
The expressions found in  \cite{Chernicoff:2008sa} for the energy and momentum of the propagating quark at finite mass diverge as the value of the external force approaches 
\beq\label{critical}
F_{crit} = \frac{2 \pi}{\sqrt{\lambda}}\ m^2.
\eeq
It is natural to interpret this in analogy with electrodynamics as the  critical value of the force capable to produce quark--antiquark pairs from the vacuum. The limiting value of the field strength can be incorporated in a non-linear generalization of electrodynamics uniquely determined by Lorentz invariance and causality proposed by Born and Infeld \cite{born}. Indeed, on the string theory side the limiting value $F_{crit}$ enters the Born-Infeld lagrangian on the D7 brane; once $F \geq F_{crit}$, the creation of open strings becomes energetically favorable, and the system becomes unstable 
\cite{Chernicoff:2008sa}. The mass $m$ in \eq{critical} should be thought of as the constituent quark mass related to the D7-brane parameter $z_m = \sqrt{\lambda} / 2\pi m$; the size of gluonic cloud around this constituent quark is $z_m$ (see \cite{Chernicoff:2008sa} and references therein).

As $F \to F_{crit}$, the energy loss of the quark jet according to \eq{radpar} and \eq{critical} is given by 
 \beq
\frac{d E_{rad}}{d t} = F_{crit}
\eeq 
and is due to the string fragmentation, i.e. creation of quark--antiquark pairs from the vacuum. The force acting on the quark thus arises from the polarization of the vacuum by the "supercritical" charge which is screened by the creation of quark--anti-quark pairs from the vacuum -- this is the mechanism of quark confinement 
proposed by Gribov \cite{Gribov:1999ui}; for a review see \cite{Dokshitzer:2004ie}. We thus have arrived at the interpretation of confinement force at zero temperature as being due to the energy loss of supercritically charged quarks in the vacuum; this interpretation provides a simple relation between the mass of the produced constituent quark $m$, the coupling $\lambda = g^2 N_c = 12 \pi \as$ in $N_c = 3$ QCD, and the string tension $\sigma$:
\beq\label{tension}
F_{crit} =   \frac{2 \pi \ m^2}{\sqrt{\lambda}} = \sigma. 
\eeq
Using $m = 300$ MeV and $\as = 0.3$ (corresponding to $\sqrt{\lambda} \simeq 3.4$), we get for the string tension $\sigma \simeq 0.85$ GeV/fm -- quite reasonable value consistent both with phenomenology and the lattice data. The corresponding size of the constituent quark is $z_m = \sqrt{\lambda} / 2\pi m \simeq 0.35 \ {\rm fm}$. Of course, ${\cal N} = 4$ SUSY Yang-Mills theory is not confining; but once confinement is introduced as an external force, the formula \eq{tension} can tell us whether this force is "critical" and would result in the production of constituent quark pairs. The numerical estimates performed above suggest that the confinement force is indeed "critical".

\medskip

The case of quark acceleration parallel to the velocity $\vec a \parallel \vec v$ applies to the fragmentation of the jets in vacuum, and to the energy loss in the direction parallel to the colliding beams in hadron collisions. However it does not apply to jets produced around mid-rapidity in nuclear collisions. High energy collisions are accompanied by the creation of strings, or longitudinal color fields, which would exert a force perpendicular to the velocity of the jet produced at mid-rapidity. Such longitudinal fields of "supercritical" strength (i.e. capable of producing quark-antiquark and gluon pairs) have been shown to emerge in heavy ion collisions  \cite{Kharzeev:2005iz,Kharzeev:2006zm,Lappi:2006fp,Fries:2006pv,Iwazaki:2007es,Fujii:2008dd,Dumitru:2008wn} from the saturated parton distributions \cite{Gribov:1984tu,Blaizot:1987nc} in the color glass condensate \cite{McLerran:1993ni}. The produced longitudinal chromo-electric and chromo-magnetic fields (termed "glasma" in \cite{Lappi:2006fp}) possess non-zero Chern-Simons number  \cite{Kharzeev:2000ef,Shuryak:2001cp,Kharzeev:2001vs,Lappi:2006fp,Janik:2002nk}; the corresponding fluctuations of the Chern-Simons number have been measured in real-time Yang-Mills calculation on the lattice \cite{Kharzeev:2001ev}.  In heavy ion collisions the presence of Chern-Simons number can induce the violation of parity \cite{Kharzeev:1998kz} (the possibility of spontaneous ${\cal T}$ and ${\cal P}$ violations has been considered in  \cite{Lee:1973iz}; in the context of heavy ion collisions, it has also been discussed by \cite{Morley:1983wr}). 
The ${\cal P}$ violation has been predicted to have a distinct signature \cite{Kharzeev:2004ey,Kharzeev:2007tn,Kharzeev:2007jp} -- charge asymmetry with respect to the reaction plane, resulting in the electric dipole moment of the produced quark-gluon matter; see \cite{Warringa:2008kv} for an overview. Recent preliminary experimental results indicate that this phenomenon may be present in RHIC data  \cite{Voloshin:2008kc}.
The propagation of charge in external classical fields of the type considered above has been considered by Shuryak and Zahed  \cite{Shuryak:2002ai}, who have evaluated synchrotron-like radiation basing on the extension of electrodynamics treatment due to Schwinger \cite{schwinger} to strong coupling; see also \cite{Zakharov:2008uk}.
\medskip 

The longitudinal color fields will result in the acceleration perpendicular to the velocity, $\vec a \perp \vec v$. In this case the formulae \eq{lien} or \eq{relat} yield the radiation power
\beq\label{radper}
\frac{d E_{rad}}{d t} = \frac{\sqrt{\lambda}}{2 \pi} \frac{1}{m^2}\ \gamma^2 \vec F^2.
\eeq 
This expression differs from \eq{radpar} in one but very important way -- it is proportional to the square of quark's energy $E = \gamma m$ (note that this will be true for any finite angle between $\vec a$ and $\vec v$).
Eq(\ref{radper}) is of course well known in classical electrodynamics \cite{Landau,Jackson} where it describes for example the energy loss due to synchrotron radiation. However, in classical electrodynamics  \eq{radper} has to be supplemented by the condition on the strength of the field so that the classical description still applies \cite{Landau,Jackson}. Namely, the field strength $G_{rest}$ (we use this notation to avoid the confusion with the force $F$) in the rest frame of the charge  must obey the inequality 
\beq\label{cond}
G_{rest} \ll \frac{m^2}{e^3}
\eeq
In the frame where the charge moves with the velocity $v$, the field strength is $G = G_{rest}/\gamma$, and the Lorentz force acting on the charge is $F = e G$. Therefore the condition \eq{cond} translates into the following condition on external force $F$:
\beq\label{condforce}
F \ll \frac{m^2}{\gamma e^2}
\eeq
or in terms of the external field $G$
\beq\label{condfield}
\gamma \frac{e^3 G}{m^2} \ll 1.
\eeq
As emphasized in \cite{Landau}, this condition does not prevent the ratio of the "radiation drag" force \eq{radper} to dominate over the external Lorentz force $F$; their ratio (we have substituted \eq{subst} for electrodynamics) is proportional to  
\beq
\frac{e^2}{m^2} \ \gamma^2\ F
\eeq
and at large energies $\gamma \gg 1$ will grow large even if \eq{condforce} is satisfied.  One is therefore justified to assume that the radiation drag force \eq{radper} is the dominant force acting on a relativistic particle at $\gamma \gg 1$. 
 
Shuryak and Zahed \cite{Shuryak:2002ai} have argued that in QCD the fields $G$ are strong, the coupling $e$ is large, and the external charge is massless, so the condition \eq{condfield} does not apply. They have thus concluded that the Li\'enard formula \eq{lien} cannot be used in QCD \cite{Shuryak:2002ai}. However, the fact that equation \eq{lien} appears as the answer in the strong coupling relativistic problem in the AdS/CFT approach encourages us to take the Li\'enard formula seriously. Indeed, the formula \eq{lien} holds in the ultra-relativistic limit; the coupling constant $\sqrt{\lambda}$ which replaces $e^2$ has been assumed large in the derivation \cite{Mikhailov:2003er}; the mass $m$ as discussed above has to be understood as a constituent mass; and the external force in AdS/CFT is limited only by the condition \eq{critical}. As emphasized above, Li\'enard formula relies only on locality, Lorentz invariance and dimensional analysis, and thus can be expected to have a wider range of validity than classical electrodynamics.

\medskip
\section{An upper bound on parton energy}

Let us now come to the point central to this paper \cite{Kharzeev:2008qr}. As was noted by Pomeranchuk  \cite{chuk,Landau}, the formula \eq{radper} has a very interesting implication\footnote{Pomeranchuk considered the radiative energy loss of cosmic ray electrons in magnetic field of Earth \cite{chuk}.}. Indeed, the power of radiation is equal to the rate of energy loss, $dE_{rad}/dt = - dE/dt = -dE/dx$, where the last equality holds for a relativistic particle with $v \simeq c$. We thus can re-write \eq{radper} as 
\beq
- \frac{dE}{dx} = \frac{\sqrt{\lambda}}{2 \pi}\ \frac{{\vec F}^2(x)}{m^4}\ E^2,
\eeq
where we have explicitly indicated the dependence of the external force $F$ on the coordinate along the path. 
This differential equation can be easily solved by noting that $dE/E^2 = - d(1/E)$, and integrating over the path of length $L$ we get
\beq
\frac{1}{E_f} = \frac{1}{E_0} + \frac{\sqrt{\lambda}}{2 \pi}\ \int_0^L dx \ \frac{{\vec F}^2(x)}{m^4}.
\eeq
As the initial energy of the quark $E_0$ increases and goes to infinity, the final energy of the parton escaping from matter tends to a constant value!

Replacing $\int dx F^2(x)$ by the product of the path length $L$ and an average value $F^2$, we get for the upper bound on the energy of the parton \cite{Kharzeev:2008qr}
\beq\label{bound}
E_{bound} = \frac{2 \pi}{\sqrt{\lambda}}\ \frac{m^4}{F^2}\ \frac{1}{L}.
\eeq
Let us measure the magnitude of the external force in units $z$ of the critical one given by \eq{critical}, $F \equiv z F_{critical}$.
We then get
\beq\label{simple}
E_{bound} = \frac{\sqrt{\lambda}}{2 \pi}\ \frac{1}{z^2 L}.
\eeq
If we assume as before in our discussion of the string tension that the external force is critical, $z \to 1$, we get from \eq{simple} a formula \cite{Kharzeev:2008qr} which depends only the value of the coupling and the path length:
\beq\label{simple1}
E_{bound} = \frac{\sqrt{\lambda_c}}{2 \pi}\ \frac{1}{L};
\eeq
where $\lambda_c$ is the critical value of the coupling corresponding to \eq{critical};
this formula applies when the masses of the propagating quark and the quarks produced from the vacuum are the same, i.e. when the propagating parton is light.

\medskip

In a more general case we can measure the magnitude of the force acting on a quark of mass $m$ in units $z_f$ of the critical value \eq{critical} for the creation of quarks of flavor $f$,
\beq
F = z^f\ F^f_{crit} = z^f\ \frac{2 \pi}{\sqrt{\lambda}}\ m_f^2; \ \ \ 0 \leq z^f \leq 1,
\eeq
the formula \eq{bound} then becomes
\beq\label{heavy}
E_{bound} = \frac{\sqrt{\lambda}}{2 \pi}\ \left(\frac{m}{m_f}\right)^4\ \frac{1}{z_f^2}\ \frac{1}{L}
\eeq

\medskip

\section{Numerical estimates}

Let us now make some numerical estimates. 
Assume first that the magnitude of the external force $F$ is given by the string tension, see \eq{tension}.  As before, we choose  
$\sqrt{\lambda} \simeq 3.4$ corresponding to $\as \simeq 0.3$, and let $z \to 1$; we then get from \eq{simple1}
\beq\label{estlight}
E_{bound}^{light} \simeq \frac{0.1}{L({\rm fm})}\ GeV;
\eeq
this means that none of the light high transverse momentum partons would escape from the longitudinal string\footnote{Note that we are discussing the case of strong coupling, and so perturbative processes are beyond the realm of this approach.}. For charm quarks, taking $m = 1.3$ GeV, $m_f = 0.3$ GeV and $z_f \to 1$ in  \eq{heavy}, we get
\beq
E_{bound}^{charm} \simeq \frac{35}{L({\rm fm})}\ GeV.
\eeq
\medskip

\begin{figure*}
\resizebox{0.9\textwidth}{!}{%
  \includegraphics{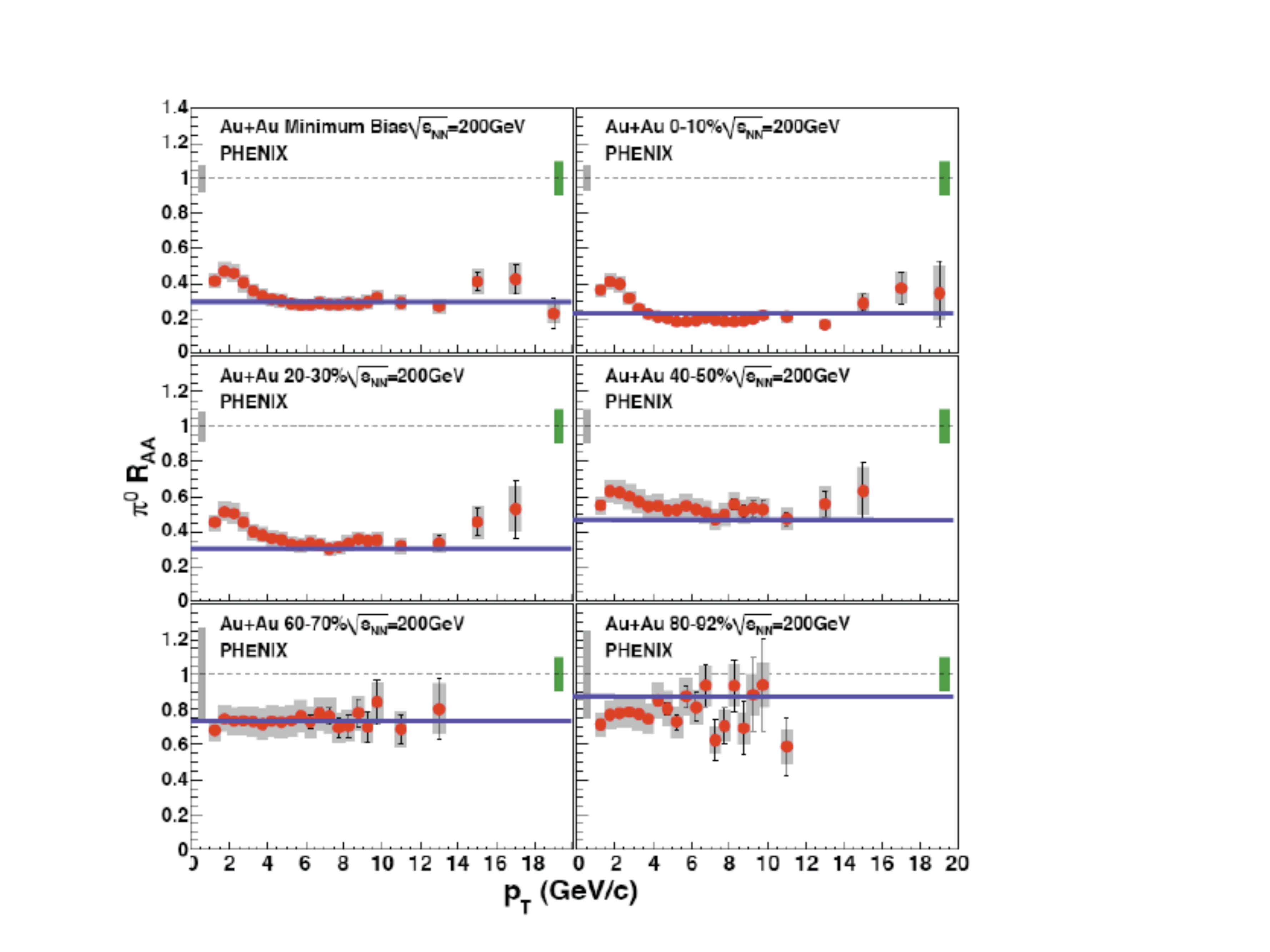}
}
\vspace*{-0.2cm}       % Give the correct figure height in cm
\caption{The data on $\pi^0$ suppression in $AuAu$ collisions at RHIC from  \cite{Adare:2008qa} compared to the universal bound model description given by Eq. \ref{ratio} (solid horizontal curves).}
\label{pizero}       % Give a unique label
\end{figure*}

\begin{figure*}
\resizebox{0.9\textwidth}{!}{%
  \includegraphics{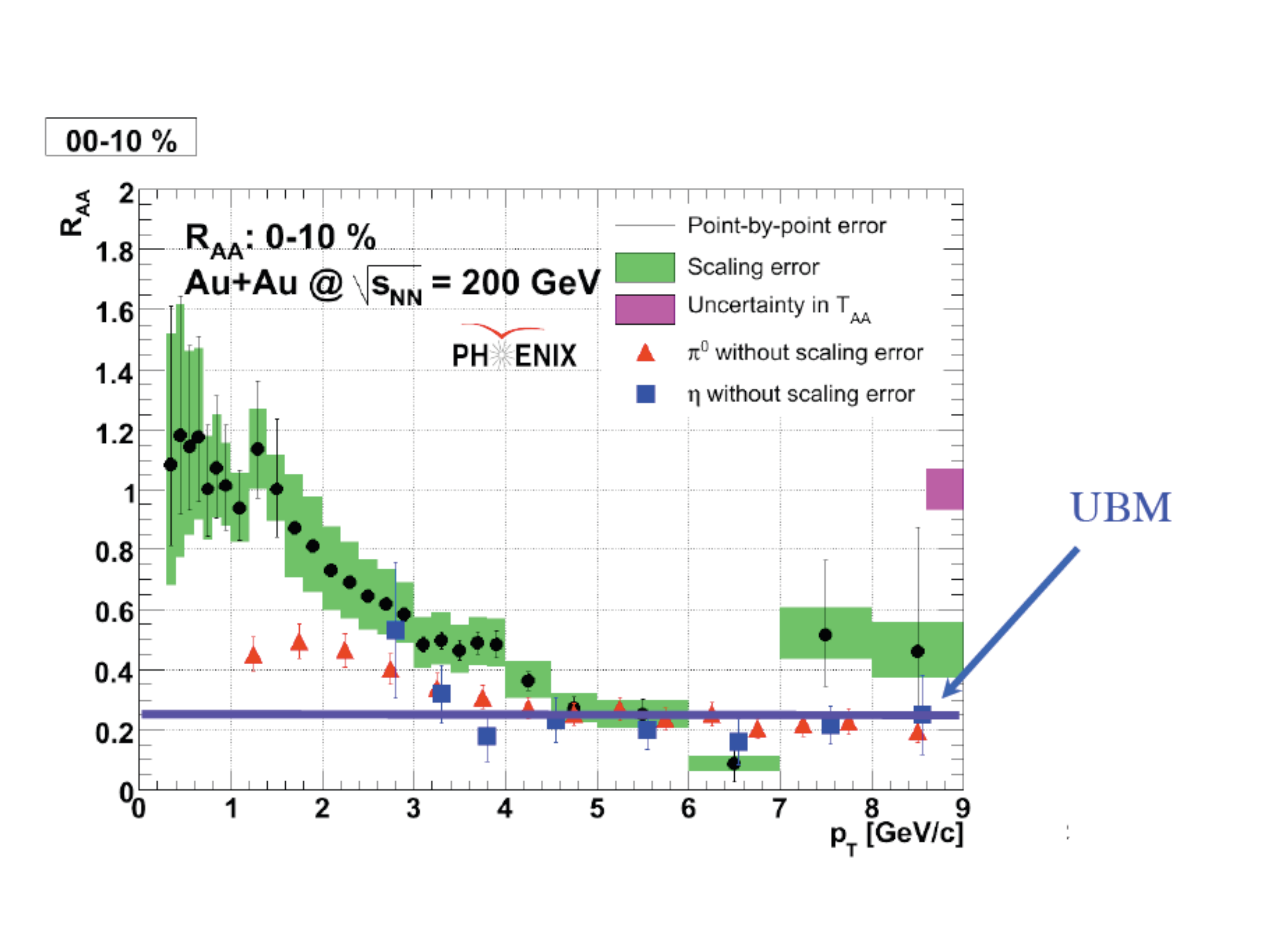}
}
\vspace*{-0.2cm}       % Give the correct figure height in cm
\caption{The data on high transverse momentum electron suppression in central $AuAu$ collisions at RHIC from \cite{Adare:2006nq} compared to the universal bound model (UBM) description given by Eq. \ref{ratio} (solid horizontal curve). High $p_T$ electrons are dominated by charm and beauty decays; also shown are the $\pi^0$ and $\eta$ data.}
\label{heavy}       % Give a unique label
\end{figure*}

Now let us take account of the fact that heavy ion collisions can produce much stronger color fields than encoded in \eq{tension}, as discussed above. It has been estimated that the magnitude of the produced fields at mid-rapidity is not much lower than the critical one needed for the production 
of charm quarks \cite{Kharzeev:2003sk}. This leads us to take $m_f = m_c$, $z_c \simeq 1$ in \eq{heavy}; we then get for charm the same estimate as we got above for light partons, \eq{estlight}
\beq
\tilde{E}_{bound}^{charm} \simeq \frac{0.1}{L({\rm fm})}\ GeV. 
\eeq
and for beauty with $m_b \simeq 4.5$ GeV
\beq\label{bquark}
\tilde{E}_{bound}^{beauty} \simeq \frac{14}{L({\rm fm})}\ GeV. 
\eeq
We should of course admit that the estimates above are very rough, and depend 
crucially on the magnitude of the produced color fields.
\medskip

These estimates suggest that in strong color fields produced in relativistic heavy ion collisions at RHIC both light and charm quarks 
and gluons cannot escape from the dense region of the produced matter. 
Of course this does not mean that there will be no high transverse momentum particles -- they will be emitted from the surface of the produced fireball, leading to a universal normalized ratio of nuclear and proton-proton cross sections $R_{AA}$ for gluon, light and charm quarks almost independent of energy. 
A weak increase of $R_{AA}$ may result from the small amount of conventional absorption in the dilute "corona" surrounding the dense core of the plasma which is expected to go away at large transverse momentum in accord with factorization theorems of perturbative QCD. This leaves little room for observing the medium-induced modifications of the jet structure -- either the jet is produced in the "corona" and is thus not modified at all, or it is produced in the dense core and is completely absorbed, with the final energy
below the bounds given above. 

Based on the perturbative arguments, 
one is led to search for the jet modification in the central collisions of heavy nuclei. The bound presented here is inversely proportional to the length of traversed medium, and so suggests that the only hope to observe 
the jet modified by the medium-induced radiation is in peripheral 
collisions and in the collisions of lighter ions. The color fields, and thus the external force acting on the color charge, are proportional to the saturation momentum squared, $F \sim Q_s^2$ which grows with the centrality as $Q_s^2 \sim N_{part}^{1/3}$ where $N_{part}$ is the number of participant nucleons (see \cite{Kharzeev:2000ph} for details). Since $L \sim N_{part}^{1/3}$ the bound may be 
expected to depend on centrality as $E_{bound} \sim 1/N_{part}$.

The color fields, and thus the external force acting on the color charge, are proportional to the saturation momentum squared $F \sim Q_s^2$ which grows with energy. Therefore  the bounds will decrease at the LHC energy. According to the estimates of the saturation momentum at the LHC and RHIC (see e.g. \cite{Kharzeev:2004if}), the bound on the $b$ quarks at the LHC will therefore change to
\beq\label{bquark}
\tilde{E}_{bound}^{beauty} \simeq \frac{2}{L({\rm fm})}\ GeV, 
\eeq 
so the suppression of $b$ quarks at high $p_{\perp}$ should become a clearly visible effect.

\section{The Universal Bound Model}\label{ubm}

To fully test the proposed universal bound, one needs to perform detailed calculations including realistic nuclear geometry and taking account of the trigger bias effect enhancing the 
contribution of the smaller path lengths. However we can try and formulate a simple model basing on the following observation. In Glauber model, a surface--to--volume ratio is quite well represented by the ratio of the number of "participants" $N_{part}$ (nucleons which underwent at least one inelastic interaction) to the number of nucleon--nucleon collisions $N_{coll}$. Indeed, 
the area of the surface scales with the linear size $R$ of the interaction region as $\sim R^2 \sim N_{part}^{2/3}$, and the volume scales as $\sim R^3 \sim N_{part}$, with the surface--to--volume ratio scaling as $ N_{part}^{-1/3}$. The number of collisions scales as $N_{coll} \sim N_{part}^{4/3}$, so to reproduce the surface--to--volume ratio we can form the ratio $N_{part}/N_{coll}$. Since as we have seen above the bound for all partons except the $b-$quarks is quite low, we can assume that all of the observed high transverse momentum particles originate entirely from the surface. The experimentally observed suppression is usually quantified through the ratio $R_{AA}$  of the cross section of high $p_T$ particle production measured in nucleus--nucleus collision to the same quantity measured in $pp$ collisions multiplied by the number of collisions $N_{coll}$; in the absence of any nuclear modifications, $R_{AA} = 1$.

This brings us to what may be called the "Universal Bound Model": 
the existence of the universal bound implies that for all high transverse momentum particles (above $p_T \simeq 5$ GeV where the collective flow effects are no longer expected to contribute) 
the ratio $R_{AA}$ is given by
\beq\label{ratio}
R_{AA} = c\ \frac{N_{part}}{2 N_{coll}}.
\eeq
Here $c \sim 1$ is an adjustable parameter of order unity; its numerical value is universal for all particles (except perhaps for hadrons containing $b$ quarks at moderate $p_T$) and is determined 
by the thickness of low-density "corona". We have introduced a factor of $2$ in the denominator of \eq{ratio} so that for $pp$ collisions $R_{pp} = c$.

The model defined by \eq{ratio} is very simple and economical, so it should be very easy to check its validity. The numbers of collisions and participants in different centrality bins can be computed within the Glauber approach (for a complete set of formulae see e.g. \cite{Kharzeev:1996yx}); we will use the tables for $N_{part}$ and $N_{coll}$ given in \cite{Kharzeev:2000ph}). We are now ready to test Eq.\ref{ratio} against the available data. Let us start with the data on light hadrons; to be specific we will consider the data on neutral pion production from \cite{Adare:2008qa}. As shown in Fig. \ref{pizero}, the model \eq{ratio} provides a reasonably good description of the data with a single parameter $c = 1.5 \pm 0.1$ which indeed appears of order unity as expected.

The next test we consider is the suppression of high $p_T$ electrons originating from the decays of charm and beauty hadrons  \cite{Adare:2006nq}. The comparison of the universal bound model (UBM) of Eq.\ref{ratio} to the data is shown in Fig. \ref{heavy}, with the parameter fixed above at $c= 1.5$. Again, we find a reasonable agreement with the data. 
 
\medskip

To summarize, if the Li\'enard formula derived through the AdS/CFT correspondence holds in the strong coupling regime in QCD then 
under quite natural assumptions about the evolution of gauge fields in heavy ion collisions there should exist an upper bound on the final energy of the parton escaping from the strongly coupled matter.  
Of course, the asymptotic freedom of QCD dictates that at some large transverse momentum the dynamics should become perturbative. However one should keep in mind that the scale relevant for the energy loss problem is not the transverse momentum of the jet, but the typical momentum transfer in the interactions of the jet with the medium. While this scale is expected to slowly increase with the transverse momentum of the jet $P_{jet}$, is it is not yet clear 
at what $P_{jet}$ the strong and weak coupling descriptions of the energy loss should match.  An experimental study of the possible existence of the bound  is needed to answer this question. 

\medskip

 This work was supported by the U.S. Department of Energy under Contract No. DE-AC02-98CH10886.

\end{document}